\documentclass[journal,twoside,web]{ieeecolor}
\usepackage{generic}
\usepackage{cite}
\usepackage{graphicx}
\usepackage{textcomp}
\usepackage{amsmath,epsfig,amssymb,url,amsfonts}
\usepackage{algorithm}
\usepackage{algpseudocode}
\usepackage{multirow}
\usepackage{url}
\usepackage{array, cite}
\usepackage{xcolor}
\usepackage{orcidlink}
\usepackage{psfrag}
\usepackage{mwe}
\usepackage{subcaption,float}
\DeclareMathOperator{\EX}{\mathbb{E}}

\newcommand{\Rbb}{{\mathbb{R}}}

\newcommand{\Ab}{{\bf A}}
\newcommand{\Lb}{{\bf L}}

\newcommand{\Tb}{{\bf T}}
\newcommand{\Hb}{{\bf H}}

\newcommand{\Rb}{{\bf R}}
\newcommand{\fb}{{\bf f}}
\newcommand{\Fb}{{\bf F}}
\newcommand{\hb}{{\bf h}}
\newcommand{\Nb}{{\bf N}}
\newcommand{\cb}{{\bf c}}

\newcommand{\Qb}{{\bf Q}}

\newcommand{\xb}{{\bf x}}

\newcommand{\yb}{{\bf y}}
\newcommand{\zb}{{\bf z}}
\newcommand{\Zb}{{\bf Z}}
\newcommand{\eb}{{\bf e}}
\newcommand{\nb}{{\bf n}}

\newcommand{\Bb}{{\bf B}}

\newcommand{\Eb}{{\bf E}}
\newcommand{\Wb}{{\bf W}}

\newcommand{\nz}[1]{\|#1\|_0} 
\newcommand{\no}[1]{\|#1\|_1} 
\newcommand{\nf}[1]{\|#1\|_F} 
\newcommand{\nt}[1]{\|#1\|_2}

\newcommand{\ninf}[1]{\|#1\|_{\infty}}
\newcommand{\Dc}{{\cal D}}

\newcommand{\Zc}{{\cal Z}}

\newcommand{\Sigb}{{\mbox{\boldmath $\Sigma$}}}
\newcommand{\phib}{{\mbox{\boldmath $\phi$}}}

\newcommand{\Sigmab}{{\mbox{\boldmath $\Sigma$}}}
\newcommand{\Phib}{{\mbox{\boldmath $\Phi$}}}
\newcommand{\thetab}{{\mbox{\boldmath $\theta$}}}
\newcommand{\omegab}{{\mbox{\boldmath $\omega$}}}
\newcommand{\argmin}{\mathop{{\rm argmin}}}

\newcommand{\supp}{\mbox{{supp}}}

\newcommand{\snr}{\mbox{{SNR}}}

\newcommand{\sbt}{\mbox{{subject~to}}}

\newcommand{\xbh}{{\mathbf{\hat x}}}

\newcommand{\Ib}{{\bf I}}

\newcommand{\Ub}{{\mathbf U}}
\newcommand{\ub}{{\mathbf u}}

\newcommand{\Vb}{{\mathbf V}}
\newcommand{\Sb}{{\mathbf S}}

\newcommand{\Nc}{{\cal N}}

\newcommand{\Pc}{{\cal P}}
\newcommand{\Qc}{{\cal Q}}
\newcommand{\Uc}{{\cal U}}

\newsavebox\mybox

\usepackage{adjustbox}
\usepackage{hyperref}
\newtheorem{theorem}{Theorem}

\newcommand{\refappendix}[1]{\hyperref[#1]{Appendix~\ref*{#1}}}
\newtheorem{lemma}[theorem]{Lemma}
\newtheorem{remark}[theorem]{Remark}
\newtheorem{definition}{Definition}
\newtheorem{assumption}{Assumption}
\begin{document}
	\title{{  Balancing Application Relevant and Sparsity Revealing Excitation in Input Design}}
	\author{Javad~Parsa\orcidlink{0000-0003-1520-4041},~Cristian R. Rojas\orcidlink{0000-0003-0355-2663},~\IEEEmembership{Member,~IEEE, } and H\r{a}kan~Hjalmarsson\orcidlink{0000-0002-9368-3079},~\IEEEmembership{Fellow,~IEEE}
		\thanks{ The authors
			are with the Division of Decision and Control Systems, School of Electrical
			Engineering and Computer Science, KTH Royal Institute of Technology, 100
			44 Stockholm, Sweden \texttt{javadp, crro, hjalmars@kth.se}}
		\thanks{This work was supported by the Swedish
			Research Council research environment NewLEADS—New Directions in
			Learning Dynamical Systems, contract 2016-06079; and Wallenberg AI,
			Autonomous Systems and Software Program (WASP), funded by Knut and
			Alice Wallenberg Foundation.}}
	\maketitle
	\begin{abstract}
	\color{black}{The maximum absolute correlation between regressors, which is called mutual coherence, plays an essential role in sparse estimation. A regressor matrix whose columns are highly correlated may result from optimal input design, since there is no constraint on the mutual coherence, making it difficult to handle sparse estimation. This paper aims to tackle this issue for fixed denominator models, which include Laguerre, Kautz, and generalized orthonormal basis function expansion models, for example.
 The paper proposes an optimal input design method where the achieved Fisher information matrix is fitted to the desired Fisher matrix, together with a coordinate transformation designed to make the regressors in the transformed coordinates have low mutual coherence. The method can be used together with any sparse estimation method and any desired Fisher matrix. A numerical study shows its potential for alleviating the problem of model order selection when used in conjunction with, for example, classical methods such as {  the Akaike Information Criterion.}}
	\end{abstract}
 \vspace{-0.25cm}
 \vspace{-0.55cm}
 \section*{Notation}
  \vspace{-0.2cm}
	{\color{black} Vectors and matrices are denoted by bold lower-case and upper-case letters, respectively. The support of a vector $\thetab$, which is denoted by supp$(\thetab)$, is the set of indices of non-zero elements in $\thetab$. $\ninf{\Bb}$ and $\nf{\Bb}$ are the maximum absolute value between all elements in matrix $\Bb$ and the Frobenius norm of the matrix $\Bb$, respectively. The vectorization operator, i.e., $\text{vec}\colon \Rbb^{a\times b}\rightarrow\Rbb^{a b}$, converts a matrix to a vector by stacking the columns	of a matrix on top of one another, and $\text{vec}^{-1}\colon \Rbb^{a b}\rightarrow\Rbb^{a\times b}$ refers to the inverse of the vectorization operator. Furthermore, $\phib_i$ denotes the i-th column of matrix $\Phib$, and $|\cdot|$ represents the absolute value.}
 \vspace{-0.45cm}
	\section{Introduction} 
  \vspace{-0.2cm}
	System identification is the science of building dynamic models of systems from data \cite{Ljung1999}. A model is an approximation of the real-life behavior of the system, and it does not provide an exact description of the system. As a consequence, different models can be found under various assumptions. A simple but widely used structure is the {  single-input single-output  linear regression model}, given by
 \vspace{-0.35cm}
	\begin{equation}
	\yb=\Phib\thetab+\eb,
	\label{eq:spsys11}
  \vspace{-0.2cm}
	\end{equation}
	where $\yb\in \Rbb^{N\times 1}$, $\thetab\in\Rbb^{n_{\theta}\times 1}$,  $\Phib\in\Rbb^{N\times n_{\theta}}$ and $\eb\in\Rbb^{N\times 1}$ denote the observations,  parameter vector, regressor matrix and noise, respectively. {\color{black} This paper considers models where $\Phib$ has a lower Toeplitz structure. This covers finite impulse response (FIR) models
	\vspace{-0.5cm}
 \begin{equation}
   y(t)=\sum_{k=1}^{n_{\theta}}\theta_k u(t-k)+e(t),\quad  t=1,\ldots,N
   \vspace{-0.27cm}  
 \end{equation}
in which case $\Phib$ is constructed from the input sequence $\{u(t)\}$. However, this structure also covers models of the form 
\vspace{-0.25cm}
\begin{equation*}
    y(t)=\sum_{k=1}^{n_{\theta}}\theta_k u'(t-k)+e(t),\quad  t=1,\ldots,N
    \vspace{-0.15cm}
\end{equation*}
    where
    \vspace{-0.35cm}
    \begin{equation}
    u'(t)=F(q)u(t)
    \label{eq:uprime}
    \vspace{-0.15cm}
\end{equation}
where 
 $F(q)$ is a fixed transfer function ($q$ is the discrete-time time-shift operator $qu(t)=u(t+1)$). In this case $\Phib$ is constructed from $\{u'(t)\}$. {  This means that the considered model covers, among others, models expressed using different shift operators such as Laguerre, Kautz, and generalized orthonormal basis function expansion models, in which case $F(q)$ has no zeros and the denominator contains the poles of the Laguerre, Kautz, or generalized basis function expansion model  \cite{293196, Heuberger&Hof&Bogsra:95}. Thus the model covers several of the most commonly used models in system identification.  We will consider the idealized case that the true system can be described by \eqref{eq:spsys11} with $\Phib$ being lower Toeplitz and $\eb\sim \Nc(0,\sigma^2 \Ib)$.}


In addition, we will consider the case where $\thetab$ is sparse, meaning that only, say, $s$ elements of this vector are non-zero, and their position is unknown. This setting appears in numerous applications. For example, in wireless communication, a common challenge is channel fading, where the received signal comprises multiple attenuated and delayed copies of the transmitted signal \cite{RePEc}. In such scenarios, the channel can be modeled using a linear FIR model with inherently sparse impulse response \cite{6034080}. Sparse estimation is a family of methods tailored to this problem, balancing the trade-off between model accuracy and complexity in system identification by finding a parameter vector with a suitable number of zero elements, which results in a model with low order and reduced complexity. A substantial literature has been published on sparse system identification, including works such as \cite{10195960, Chiuso2012}. For example, in \cite{Chiuso2012}, a Bayesian approach for sparse system identification is proposed, while the parameters of a linear regression model are estimated using a coordinate transformation in \cite{10195960}. It is also worth pointing out that there may be other prior information present than that the parameter vector is sparse. A very common problem is that of model order selection, in which case it is known that the first elements of the parameter vector are non-zero but the exact number of non-zero elements is not known. We will illustrate in Section \ref{sec:nums} that also this problem is alleviated by the proposed method. 

The mutual coherence of the regressor matrix $\Phib$, defined as 
 \vspace{-0.3cm}
	\begin{equation}
	\mu_{\Phib}=\max_{i\neq j}\frac{|\langle\phib_i,\phib_j\rangle|}{\nt{\phib_i}\nt{\phib_j}},
	\label{eq:muphi}
 \vspace{-0.2cm}
	\end{equation}
 where $\{\phib_j\}$ are the columns of $\Phib$, 
plays an important role in sparse estimation, as evidenced by extensive publications on the subject, e.g. \cite{Elad10, Ben-Haim2010, 8553417}. These works consistently highlight the importance of maintaining low mutual coherence to achieve accurate sparse estimation.
Now, in system identification, the columns of the regressor matrix in  \eqref{eq:spsys11} are typically taken as time-shifted versions of the same signal, e.g. the input. When this signal has strong temporal correlation, this leads to a high mutual coherence, cf. the definition \eqref{eq:muphi}. This problem may be exacerbated by optimal input design which often results in an input that is tuned to rather narrow frequency bands, implying high temporal correlation, aimed at revealing system properties of importance for the intended model use. In Section \ref{sec:nums} this is illustrated for a control design problem where a frequency band covering the cross-over frequency is excited.

Optimal input design typically shapes the Fisher Information Matrix (FIM) using the input signal as a decision variable to meet some design objectives. In model \eqref{eq:spsys11} with $\eb\sim \Nc(0,\sigma^2\Ib)$, the FIM is given by\footnote{We will not need the Gaussian assumption in the following but, with some misuse of notation, will continue to use the term FIM for $\Ib_F$, even though without the Gaussian noise assumption, $\Ib_F$ is not the FIM but the inverse covariance matrix of the least-squares estimate of $\thetab$ in the model \eqref{eq:spsys11} which can also be used in input design.}:
\vspace{-0.25cm}
\begin{equation}
\Ib_F = \frac{1}{\sigma^2}\Phib^T\Phib.
\label{eq:filinreg}
\vspace{-0.25cm}
\end{equation}
This implies that the mutual coherence $\mu_\Phib$ in \eqref{eq:muphi} is directly influenced by the Fisher matrix. {  The purpose of this contribution is to, starting from a desired Fisher information matrix, typically, but not necessarily, obtained by solving an optimal input design problem, develop a method that generates an input such that this Fisher information matrix is approximately achieved but also limits the mutual coherence so as to facilitate sparse estimation via a linear variable transformation. Different from the Frequency Domain Method (FDM) method, the input is not obtained by way of spectral factorization and filtering white noise. Instead, the input sequence is directly designed.} A previous study \cite{9838242} introduced a method for input design based on a linear coordinate transformation with two terms. This method employs an approximation of mutual coherence using the Frobenius norm. {  Although this method utilized convex optimization with three sub-steps, it was found to be slow using generic software such as CVX, particularly for large-scale problems where the number of parameters and observations in \eqref{eq:spsys11} is on the order of a hundred.} Additionally, the use of the Frobenius norm approximation was not ideal, focusing on reducing correlation between rows rather than columns \cite{8553417}.
 \vspace{-0.51cm}
\subsection*{Contribution}
\vspace{-0.2cm}
{\color{black}As a remedy to the shortcomings in \cite{9838242}, we propose an alternative approach with a linear coordinate transformation that instead uses the exact formulation of mutual coherence. According to Section \ref{sec:nums}, exact formulation results in a significant reduction in mutual coherence and subsequently lower error compared to \cite{9838242}. Also the proposed method provides a closed-form solution in each step that significantly reduces the computation time, as evidenced in Section \ref{sec:nums}. Additionally, our proposed approach is versatile, capable of handling commonly used linear models in system identification, such as FIR, Laguerre, Kautz, and generalized orthonormal basis function expansion models, in contrast to the method presented in \cite{9838242}, which is specifically designed for the FIR model. It is worth noting that the proposed method can be combined with any standard sparse estimation algorithm. In summary, the contributions of
our paper are:
\vspace{-0.2cm}
	\begin{enumerate}
		\item A new optimal input design criterion suitable for sparse estimation problems in the linear regression model \eqref{eq:spsys11} is proposed. The method combines a coordinate transformation of the model \eqref{eq:spsys11} with {  estimation accuracy} and low mutual coherence constraints.
  \item A numerical algorithm is proposed to solve the optimization problem. The method is a three step iterative procedure using proximal mappings and alternating minimization. In comparison with \cite{9838242}, the algorithm can handle fixed denominator types of models (see above), and is computationally much more efficient. 
		\item Similar to \cite{9838242}, the proposed method is numerically evaluated in a  model reference control problem where the actual model order is unknown. However, in addition to its use together with standard sparse algorithms, as done in \cite{9838242}, it is also combined with the classical {  AIC (Akaike Information Criterion) and BIC (Bayesian Information Criterion)} model order selection methods. In addition, the proposed method is compared with the method in \cite{9838242}, both in terms of accuracy and computation time.  
	\end{enumerate} }
 \vspace{-0.65cm}
	\subsection*{Structure of the Paper}
	\vspace{-0.2cm} 
	The remaining parts of the paper can be summarized as follows: In the next section, we give a brief recap of sparse estimation and optimal input design. The novel optimal input design, incorporating low mutual coherence, is formulated as an optimization problem in Section \ref{sec:prop1}. In Section \ref{sec:prop2}, an iterative algorithm is proposed for solving the optimization problem. In the numerical study part, Section \ref{sec:nums}, the proposed method is evaluated against state-of-the-art algorithms.
 \vspace{-0.5cm}
\section{Recap of sparse estimation and optimal input design}
 \vspace{-0.3cm}
Before proceeding with the core of the paper, a brief review of sparse estimation and optimal input design is in place.
\vspace{-0.55cm} 
\subsection{Sparse estimation}
\vspace{-0.2cm} 
{\color{black} To estimate a sparse parameter vector $\thetab$ in the model \eqref{eq:spsys11}, the error $\nt{\yb-\Phib\thetab}^2$ is minimized with respect $\thetab$ while the $\ell_1$ norm of $\thetab$, $\no{\thetab}$, is constrained to be smaller than $\tau$ for achieving the desired sparsity. This method is known as LASSO (least absolute shrinkage and selection operator), and will be denoted by $P_1^\tau$ in the following. Greedy methods, such as Orthogonal Matching Pursuit (OMP) \cite{PatiRK93}, and Generalized OMP (GOMP) \cite{WangKwonShim2012}, is another family of sparse estimation algorithms. }
\vspace{-0.6cm}
\subsection{Optimal input design}\label{subsec:optimalinp}
\vspace{-0.15cm}
 Experimental data are noisy, resulting in uncertainty in the \\\\estimated model. Input design may be utilized to improve the accuracy of an estimator so that desired criteria are met.
 Many different methods exist, and we refer to  \cite{Goodwin1977, Ljung1999, Bombois2006, Hjalmarsson2009} and references therein for a wide range of examples. As for linear time-invariant systems the per sample Fisher matrix is linear in the input spectrum, many methods proceed in two steps: i) The optimal input spectrum, together with the optimal FIM $\Ib_F^d$, is determined using (convex) optimization, 
 ii) filtering a realization of white noise through a stable minimum phase spectral factor of the optimal input spectrum.  This approach is called the FDM in the following. 
 Least-costly identification \cite{Bombois2006} and application-oriented input design \cite{Hjalmarsson2009} are examples of methods of this type. They
 minimize experimental effort, such as input power, subject to quality constraints related to the intended use of the model, where these constraints are expressed in terms of the per sample FIM. We will employ this method in the numerical example in Section \ref{sec:nums} where more details will be provided. }
 \vspace{-0.45cm}
	\section{An input design criterion incorporating mutual coherence }\label{sec:prop1}
	\vspace{-0.2cm} 
	{\color{black} The following will be our standing assumption throughout the paper.
	 \begin{assumption}
	     The data obeys model \eqref{eq:spsys11}, with $\thetab=\thetab_0$ denoting the true model parameter. It holds that ${\rm E}[\eb]=0$, ${\rm E}[\eb\eb^T]=\sigma^2 \Ib$, $N\geq n_\theta$, $\Phib^T\Phib>0$ with $\Phib$ being lower Toeplitz. 
	 \end{assumption}}
 {\color{black}
 As outlined in Section \ref{subsec:optimalinp}, many optimal input design methods in the first step determine the optimal input spectrum and the desired FIM $\Ib_F^d$. Alternatively to realizing the input by way of filtering (see Section \ref{subsec:optimalinp}), one may obtain the optimal regressor by minimizing the difference between the two sides in \eqref{eq:filinreg}, with $\Ib_F=\Ib_F^d$. In \cite{EBADAT20141422} the Frobenius norm is used to this end. Consequently,
 \vspace{-0.25cm} 
	\begin{equation}
	\min_{\Phib\in\Pc} \nf{\Phib^T\Phib-\sigma^2\Ib_F^d}^2,
	\label{eq:inf1}
	\vspace{-0.2cm} 
	\end{equation} 
in which $\Pc$ denotes the set of $N\times n_{\theta}$ Toeplitz matrices due to the considered linear FIR model in \cite{EBADAT20141422}.
The achieved regressor from \eqref{eq:inf1} can then be used in the identification experiment. An intriguing question arises: at what level of mutual coherence for the derived regressor from \eqref{eq:inf1} does it lead to accurate sparse estimation of $\thetab$? The following result addresses this question.
 \begin{theorem}[Stability of $P_1^{\tau}$ \cite{Ben-Haim2010}] \label{thm:5}
Consider the model \eqref{eq:spsys11}, where $\eb\sim\Nc(0, \sigma^2\Ib)$. Here, $\thetab_0$ denotes a parameter vector with sparsity characterized by the support set $\Lambda_0$, and $\nz{\thetab_0} = s$. Assume that the mutual coherence of the matrix $\Phib$ satisfies the condition $0 \leq \mu_{\Phib} < \frac{1}{3s}$. Thus,
under these conditions, with a probability of at least
\vspace{-0.2cm}
\begin{equation}
\Big(1 - \frac{(n_{\theta} - s)}{n_{\theta}^{1+\nu}}\Big)\Big(1 - e^{-\frac{s}{7}}\Big), \quad \forall \nu > 0,
\label{eq:prog}
\vspace{-0.2cm}
\end{equation}
the support set of the solution ${\thetab}^{\epsilon}_1$ obtained through $P_1^{\tau}$ is $\Lambda_0$.
\label{thm:l1normsp}
\end{theorem}
\vspace{-0.3cm}
According to Theorem \ref{thm:l1normsp}, accurate sparse estimation of parameter vector is achievable using $P_1^{\tau}$ if there is a small mutual coherence.  However, since there is no mutual coherence constraint in \eqref{eq:prog}, its optimal solution often does not meet the conditions of the above theorem. This is evident from the optimal regressor in the numerical study of Section \ref{sec:nums}.
 
 To incorporate the need for low mutual coherence, we borrow the idea of employing a linear coordinate transformation from \cite{10195960}, and write the model \eqref{eq:spsys11} as
 \vspace{-0.2cm}
	\begin{align}
	\left\{
	\begin{array}{lr}
	\yb = \Phib\Hb^{-1}\xb+\eb\\
	\xb = \Hb\thetab,
	\end{array}
	\label{eq:prometh}\right.
	\end{align}
	in which the matrix $\Hb$ is utilized to transform the coordinates such that the sparse parameter vector $\thetab$ is estimated in the new coordinates $\xb$. In \eqref{eq:prometh} the vector $\xb$ is not necessarily sparse. Hence, we can use least squares to estimate this vector as follows:
 \vspace{-0.32cm}
	\begin{equation}
	\xbh=\argmin_{\xb}\nt{\yb-\Phib\Hb^{-1}\xb}^2.
	\label{eq:ls}
	\vspace{-0.22cm}
	\end{equation}
	Solving the above optimization problem leads to
 \vspace{-0.2cm}
	\begin{align}
	\xbh = \Hb(\Phib^T\Phib)^{-1}\Hb^T(\Phib\Hb^{-1})^{T}\yb = \xb + \omegab = \Hb\thetab + \omegab,
	\label{eq:newmod}
	\end{align}
	where $\omegab = \Hb(\Phib^T\Phib)^{-1}\Hb^T(\Phib\Hb^{-1})^{T}\eb$ and $\EX[\omegab\omegab^T]=\sigma^2\Hb(\Phib^T\Phib)^{-1}\Hb^T$. In the new linear regression model \eqref{eq:newmod}, the parameter vector $\thetab$ is estimated by using the regressor $\Hb$. Hence, this new regressor should have a low mutual coherence for accurately estimating the parameters when the vector $\thetab$ is sparse. On the other hand, the coordinate transformation has an undesired effect on the noise since the covariance matrix of $\omegab$ in \eqref{eq:newmod} is not an identity matrix. The following lemma provides an expression for the mutual coherence of $\Hb$.
 \begin{lemma}[\cite{10195960}]
    Consider the matrix $\Hb$ with columns normalized to unit length. The mutual coherence of $\Hb$, denoted as $\mu_{\Hb}$, then is given by
    \vspace{-0.3cm}
    \begin{equation}
       \mu_{\Hb} = \ninf{\Hb^{T}\Hb-\Ib}.
    \end{equation}
    \label{lem:mutro}
    \vspace{-0.6cm}
 \end{lemma}
 
 In \cite{10195960} where the purpose was to design $\Hb$ for facilitating sparse estimation without input design, to find the optimal $\Hb$, the following optimization problem was proposed: 
 \vspace{-0.22cm}
	\begin{align}
	\nonumber&\min_{\Hb\in\Dc}\ninf{\Hb(\Phib^T\Phib)^{-1}\Hb^T-\Ib}\\&
	\text{subject to}~~\ninf{\Hb^{T}\Hb-\Ib}\leq\alpha_1,
	\label{eq:mainpr1}
 \vspace{-0.2cm}
	\end{align}
	where $\Dc = \{\Hb \in \mathbb{R}^{n_{\theta} \times n_{\theta}}: \nt{\Hb(:,i)}=1,\, \forall~ 1\leq i\leq n_{\theta}\}$.
	In the above problem, the cost function is used to minimize the undesired effect of the coordinate transformation on the noise. Simultaneously, the constraint, as indicated by Lemma \ref{lem:mutro}, guarantees that the newly transformed regressor exhibits low mutual coherence provided $\alpha_1$ is chosen small enough.
 
	In \eqref{eq:mainpr1}, the cost function minimizes the difference between $\Hb(\Phib^T\Phib)^{-1}\Hb^T$ and the identity matrix. Using the Frobenius norm instead of the infinity norm simplifies the process of solving the optimization problem. Consequently, here we propose to instead use}
	\vspace{-0.3cm}
	\begin{align}
	\nonumber&\min_{\Hb\in\Dc}\nf{\Hb(\Phib^T\Phib)^{-1}\Hb^T-\Ib}^2\\&
	\text{subject to}~~\ninf{\Hb^{T}\Hb-\Ib}\leq\alpha_1.
	\label{eq:mainpr11}
	\vspace{-0.3cm}
	\end{align}
	Finally, to incorporate also the degrees of freedom offered by input design, we combine \eqref{eq:inf1} with \eqref{eq:mainpr11}. 
	 This results in the following optimization problem, where the coordinate transformation and input are jointly designed to meet the objectives of low mutual coherence and achieving the optimal FIM: 
 \vspace{-0.5cm}
	\begin{align}
	\nonumber\min_{\Hb\in\Dc, \Phib\in\Pc}&\nf{\Hb(\Phib^T\Phib)^{-1}\Hb^T-\Ib}^2\\
	\text{subject to}~~~\nonumber&\ninf{\Hb^{T}\Hb-\Ib}\leq\alpha_1\\
	&\nf{\Phib^T\Phib-\Ib_F^d}^2\leq\alpha_2.
	\label{eq:conspro}
	\end{align}
{\color{black} In \eqref{eq:inf1}, the constraint $\Phib\in\Dc$ was utilized to enforce
the structure of a linear FIR model.
From $\Phib$ we can recover the sequence $u'(t)$, $t=1-n_\theta,\ldots,N-1$. However, unless $F(q)=1$, i.e. a FIR model is used, an additional step has to be employed to obtain the input $u(t)$. From \eqref{eq:uprime} it follows that
\vspace{-0.35cm} 
\begin{equation}
u(t)=F^{-1}(q)u^{\prime}(t).
\vspace{-0.2cm} 
\end{equation}
Notice that in the case that $F(q)$ has zeros outside the unit circle, the filtering needs to be non-causal to enforce stability.}
 \vspace{-0.9cm}
	\section{Algorithm}\label{sec:prop2}
 \vspace{-0.1cm}
{\color{black} The optimization problem \eqref{eq:conspro} is non-convex and in this section, we will outline an iterative numerical method for solving it. Each iteration consists of three steps using alternating minimization where the first step is based on the proximal mapping of the infinity norm. We begin by transforming \eqref{eq:conspro} into an unconstrained version by adding the constraints as penalties to the cost function, resulting in the Lagrangian}
  \vspace{-0.2cm}
		\begin{align}
		\nonumber\min_{\Hb\in\Dc, \Phib\in\Pc}&\nf{\Hb(\Phib^T\Phib)^{-1}\Hb^T-\Ib}^2+\lambda\ninf{\Hb^T\Hb-\Ib}\\&+\lambda^{\prime}\nf{\Phib^T\Phib-\Ib_F^d}^2,
		\label{eq:mainpr}	
  \vspace{-0.2cm}
		\end{align}
where $\lambda$ and $\lambda^{\prime}$ denote positive penalty parameters. The penalty parameters can be adjusted in an iterative procedure so that the constraints in \eqref{eq:conspro} are met. To solve the optimization problem \eqref{eq:mainpr} we define the new variable $\Nb =\Hb^{T}\Hb-\Ib$, which converts this problem to the constrained problem
\vspace{-0.15cm}
	\begin{align*}
	\begin{array}{cl}
	\min\limits_{\Phib\in\Pc, \Nb\in\Nc, \Hb\in\Dc}  &\nf{\Hb(\Phib^T\Phib)^{-1}\Hb^T-\Ib}^2+\lambda^{\prime}\nf{\Phib^T\Phib-\Ib_F^d}^2\\&+\lambda\ninf{\Nb}\\
	\text{subject to }&\Nb =\Hb^{T}\Hb-\Ib,
	\end{array}
 \vspace{-0.4cm}
	\end{align*}
	in which $\Nc$ indicates that $\Nb$ is constrained to be a a symmetric matrix for which $\Nb+\Ib>0$.  
	We proceed and add also the new constraint to the cost function, resulting in  
 \vspace{-0.2cm}
	\begin{align}
	\nonumber\min\limits_{\Phib\in\Pc, \Nb\in\Nc, \Hb\in\Dc}&  \nf{\Hb(\Phib^T\Phib)^{-1}\Hb^T-\Ib}^2+\lambda^{\prime}\nf{\Phib^T\Phib-\Ib_F^d}^2\\&+\lambda\ninf{\Nb}+\frac{1}{2\varsigma}\nf{\Nb+\Ib-\Hb^{T}\Hb}^2,
	\label{eq:alter1}
 \vspace{-0.2cm}
	\end{align} 
	where $\varsigma$ is a positive penalty parameter.
	In the above optimization problem, the three matrices, $\Nb$, $\Hb$, and $\Phib$, act as decision variables. Next, alternating minimization with three steps is used, where in each step one of these three matrices is optimized, while the other variables are kept fixed. This gives the following procedure:
	
	1) Update $\Nb$:
 \vspace{-0.3cm}
	\begin{equation}
	\Nb_{k+1}^c=\argmin_{\Nb\in\Nc}\lambda\ninf{\Nb}+\frac{1}{2\varsigma}\nf{\Nb+\Ib-\Hb_k^{T}\Hb_k}^2.
	\label{eq:step1}
	\end{equation} 
	
	2) Update $\Hb$:
 \vspace{-0.3cm}
	\begin{align}
	\Hb_{k+1}=\argmin_{\Hb\in\Dc}\, &\nf{\Hb(\Phib^T_k\Phib_k)^{-1}\Hb^T-\Ib}^2 \nonumber \\
	&+\frac{1}{2\varsigma}\nf{\Nb_{k+1}+\Ib-\Hb^{T}\Hb}^2.
	\label{eq:step2}
	\end{align}
	
	3) Update $\Phib$:
 \vspace{-0.3cm}
	\begin{align}
	\nonumber\Phib_{k+1}=\argmin_{\Phib\in\Pc}&\nf{\Hb_{k+1}(\Phib^T\Phib)^{-1}\Hb^T_{k+1}-\Ib}^2 \\&+\lambda^{\prime}\nf{\Phib^T\Phib-\Ib_F^d}^2.
	\label{eq:step3}
	\end{align} 
We use proximal mapping in the first step because there is a closed-form solution for the proximal mapping of the infinity norm. We minimize the cost function \eqref{eq:step1} without the constraint $\Nc$, and then the solution is projected into $\Nc$.   
	\begin{definition}[\cite{PariB14}]
		Let $p\colon \text{dom}_{p} \rightarrow(-\infty,+\infty]$ be a proper and lower semicontinuous function. Then the proximal mapping of this function at $\xb\in \Rbb^{n}$ is defined as
		\vspace{-0.3cm}
		\begin{equation}
		\text{prox}_{p}(\xb)=\argmin_{\ub\in \text{dom}_p}\left\{\frac{1}{2}\nt{\xb-\ub}^2+p(\ub)\right\}.
		\label{eq:proximal}
		\end{equation}
	\end{definition}

	Since in the definition of proximal mapping, we deal with vectors, the vectorization operator vec is applied to the first step of the above alternating minimization, which gives
	\vspace{-0.3cm}
	\begin{equation}
	\nb_{k+1}=\argmin_{\nb}\lambda\ninf{\nb}+\frac{1}{2\varsigma} \nt{\nb -\fb_{k}}^2,
	\label{eq:findma3}
	\vspace{-0.3cm}
	\end{equation}
	in which $\nb\triangleq\text{vec}(\Nb)$ and $\fb_{k}\triangleq\text{vec}(\Fb_k)$, with $\Fb_k \triangleq \Hb_k^{T}\Hb_k-\Ib$. According to the definition of proximal mapping in \eqref{eq:proximal}, the above problem is the proximal mapping of the function $\lambda~\varsigma\ninf{.}$ at $\fb_{k}$. Hence, we can write 
  \vspace{-0.3cm}
	\begin{equation}
	\nb_{k+1} = \text{prox}_{\lambda~\varsigma\ninf{.}}(\fb_{k}).
	\label{eq:proxma}
  \vspace{-0.2cm}
	\end{equation}
	Due to the presence of the infinity norm, a direct computation of this proximal mapping may be difficult. However, the following lemma is of help.
	
	\begin{lemma} [Moreau decomposition~\cite{PariB14}]
		For a proper and lower semicontinuous function $p\colon\Rbb^m \rightarrow(-\infty,+\infty]$ whose conjugate is denoted by $p^*$, the following relation holds at $\hb\in\Rbb^m$ and $\eta>0$:
   \vspace{-0.4cm}
		\begin{equation}
		\hb = \text{prox}_{\eta p}(\hb)+\eta\,\text{prox}_{\frac{p^*}{\eta}}\left(\frac{\hb}{\eta}\right).
		\label{eq:lemmamo}
		\end{equation}
	\end{lemma}
 
    Assuming that $p = \ninf{.}$ and $\eta = \lambda~\varsigma$ in \eqref{eq:lemmamo}, and noting that the conjugate function of $\ninf{.}$ is the $\ell_1$-norm, the proximal mapping in \eqref{eq:proxma} can be computed as
    \vspace{-0.2cm}
	\begin{equation}
	\text{prox}_{\lambda~\varsigma\ninf{.}}(\fb_k)=\fb_k-\lambda~\varsigma\,\text{prox}_{\frac{\no{.}}{\lambda~\varsigma}}\left(\frac{\fb_k}{\lambda~\varsigma}\right).
	\label{eq:fiprox1}
 \vspace{-0.1cm}
	\end{equation}
	Furthermore, it is straightforward to show that
  \vspace{-0.2cm}
	\begin{equation}
	\text{prox}_{\frac{\no{.}}{\lambda~\varsigma}}\left(\frac{\fb_k}{\lambda~\varsigma}\right)=\frac{1}{\lambda~\varsigma}\Pc_{\Dc_2}(\fb_k),
	\label{eq:proxma1}
  \vspace{-0.2cm}
	\end{equation}
	where $\Dc_2$  denotes the $\ell_1$-norm ball with radius $\lambda~\varsigma$, and $\Pc_{\Dc_2}$ denotes the orthogonal projection onto $\Dc_2$.
	Using \eqref{eq:proxma1} into \eqref{eq:fiprox1} yields 
  \vspace{-0.45cm}
	\begin{equation}
	\text{prox}_{\lambda~\varsigma\ninf{.}}(\fb_k)=\fb_k-\Pc_{\Dc_2}(\fb_k).
	\label{eq:finalvec}
  \vspace{-0.15cm}
	\end{equation}
	Next the inverse of the vectorization operator $\text{vec}^{-1}$ is applied to \eqref{eq:finalvec}, resulting in
  \vspace{-0.2cm}
	\begin{equation}
	\Nb_{k+1}=\text{vec}^{-1}(\text{prox}_{\lambda~\varsigma\ninf{.}}(\fb_k))=\Fb_k-\text{vec}^{-1}(\Pc_{\Dc_2}(\fb_k)).
	\label{eq:finalvec1}
	\end{equation}
	Next the above solution is projected into $\Nc$. Since in \eqref{eq:finalvec1} the projection $\Pc_{D_2}$ operates element-wise and the matrix $\Fb_k$ is symmetric, $\Nb_{k+1}$ is symmetric. To enforce $\Nb+\Ib>0$, we set 
 \vspace{-0.5cm}
	\begin{equation}
	\Nb_{k+1}^c =\Ab^+-\Ib 
	\label{eq:podema}
 \vspace{-0.2cm}
	\end{equation} 
	where $\Nb_{k+1}+\Ib = \Eb\Lb\Eb^T$ denotes the SVD (Singular Value Decomposition) of $\Nb_{k+1}+\Ib$, $\Ab^+ = \Eb\Lb^+\Eb^T$  and $\Lb^+ = \max(\epsilon\Ib,\Lb)$ in which the $\max$ operator takes the element-wise maximum between the diagonal entries of two matrices and $\epsilon$ is a very small positive real value. This gives the nearest positive definite matrix to $\Nb_{k+1}$  in the Frobenius norm sense.
	 
	To summarize by using proximal mapping, a closed-form solution to update the matrix $\Nb_{k+1}^c$ in the first step of cyclic minimization was found.
	
In the second and third steps of the alternating minimization, the cost functions  \eqref{eq:step2} and \eqref{eq:step3} are  minimized with respect to $\Hb$ and $\Phib$, respectively, while the matrix $\Nb_{k+1}$ is known from \eqref{eq:finalvec1}. Although the main challenge in the optimization problem \eqref{eq:alter1} with respect to $\Hb$ and $\Phib$ is that the problem is non-convex, the class of unconstrained matrices when the cost function is zero satisfies (see~\cite{Stoica2008})
 \vspace{-0.2cm}
	\begin{equation*}
	\Hb(\Phib^T\Phib)^{-1}\Hb^T = \Ib, \Hb^{T}\Hb = \Nb_{k+1}^c+\Ib,~\text{and}~
	\Phib^{T}\Phib = \Ib_F^d.
  \vspace{-0.1cm}
	\end{equation*}
	The above equations are equivalent to the formulas
  \vspace{-0.1cm}
	\begin{equation}
	\Phib = \Ub\Hb,
	\Hb = \Qb(\Nb_{k+1}^c+\Ib)^{\frac{1}{2}}, ~\text{and}~
	\Phib = \Zb(\Ib_F^d)^{\frac{1}{2}},
	\label{eq:finalvec1112}
  \vspace{-0.1cm}
	\end{equation}
	where $\Ub\in\Rbb^{N\times n_{\theta}}$ and $\Zb\in\Rbb^{N\times n_{\theta}}$ are semi-unitary matrices while $\Qb\in\Rbb^{n_{\theta}\times n_{\theta}}$  is a unitary matrix. Thus, the optimization problems~\eqref{eq:step2} and \eqref{eq:step3} can be relaxed to
	
	2) Update $\Hb$:
 \vspace{-0.2cm}
		\begin{align}
		\nonumber &(\Hb_{k+1},\Qb_{opt},\Ub_{opt})=\\&  \argmin_{\Hb\in\Dc, \Qb\in\Qc, \Ub\in\Uc}\,\nf{\Ub\Hb-  \Phib_k}^2
		+\frac{1}{\varsigma}\nf{\Hb- \Qb(\Nb_{k+1}^c+\Ib)^{\frac{1}{2}}}^2.
		\label{eq:step22}
  \vspace{-0.2cm}
		\end{align}
		
		3) Update $\Phib$:
  \vspace{-0.2cm}
			\begin{align}
			&(\Phib_{k+1},\Ub_{opt},\Zb_{opt}) \nonumber\\
			&\quad =\argmin_{\Phib\in\Pc,\Ub\in\Uc,\Zb\in\Zc} \nf{\Phib- \Ub\Hb_{k+1}}^2 +\lambda^{\prime}\nf{\Phib- \Zb(\Ib_F^d)^{\frac{1}{2}}}^2,
			\label{eq:step3p}
			\end{align}
	where $\Uc$, $\Qc$ and $\Zc$ are defined as $\Uc=\{\Ub\in\Rbb^{N\times n_{\theta}}\colon \Ub^T\Ub=\Ib\}$, $\Qc=\{\Qb\in\Rbb^{n_{\theta}\times n_{\theta}}\colon \Qb^T\Qb=\Ib\}$ and $\Zc=\{\Zb\in\Rbb^{N\times n_{\theta}}\colon \Zb^T\Zb=\Ib\}$, respectively.
	
	To solve the optimization problem \eqref{eq:step22}, we have three variables. Hence, to find these variables, the following  cyclic minimization with two sub-steps is proposed: 
	
	Step 2.1: The semi-unitary matrices $\Ub$ and $\Qb$ are kept fixed to their most recent values, and the problem \eqref{eq:step22} is solved with respect to $\Hb$:
 \vspace{-0.25cm}
	\begin{equation*}
	\min_{\Hb\in\Dc}\nf{\Ub_{opt}\Hb-  \Phib_k}^2+\frac{1}{\varsigma}\nf{\Hb- \Qb_{opt}(\Nb_{k+1}^c+\Ib)^{\frac{1}{2}}}^2.
 \vspace{-0.1cm}
	\end{equation*} 
	To solve the above minimization problem, the first derivative of this cost function should be zero, which gives:
  \vspace{-0.3cm}
	\begin{align}
	\Hb_{k+1} = \big(1+\frac{1}{\varsigma}\big)^{-1} \bigg(\Ub_{opt}^T\Phib_{k} +\frac{1}{\varsigma}\Qb_{opt}(\Nb_{k+1}^c+\Ib)^{\frac{1}{2}}\bigg).
	\label{eq:udateh}
	\end{align} 
	Due to the restriction to $\Dc$, each column of $\Hb_{k+1}$ should be normalized, i.e.,
 \vspace{-0.4cm}
	\begin{equation}
	\Hb_{k+1}(:,i) = \frac{\Hb_{k+1}(:,i)}{\nt{\Hb_{k+1}(:,i)}}, \; \forall~ 1\leq i \leq n_{\theta}.
	\label{eq:normali}
	\end{equation}
	
	Step 2.2: The
	optimal $\Ub$ and $\Qb$ in \eqref{eq:step22} can be found through SVD when $\Hb$ is known from \eqref{eq:udateh}, i.e.,
	\begin{align}
	\nonumber&{\Hb_{k+1}\Phib_k^T}=\bar{ \Ub}\Sigb{\tilde{\Ub}}^T, \Ub_{opt}={\tilde{\Ub}}\bar{\Ub}^T, \\
	&{\Hb^T_{k+1}}(\Nb_{k+1}^c+\Ib)^{\frac{1}{2}}=\bar{\Vb}\tilde{\Sigb}{\tilde{\Vb}}^T, \Qb_{opt}=\bar{\Vb}{\tilde{\Vb}}^T.
	\label{eq:updateuq}
	\end{align}
	
Finally, in the third step of the cyclic algorithm, i.e. the solution of \eqref{eq:step3}, the cost function \eqref{eq:step3p} is minimized with respect to $\Phib$, $\Zb$ and $\Ub$. 
	Now, each Toeplitz matrix, e.g., $\Phib$, can be written as $\Phib = \sum\limits_{l=-N+1}^{n_{\theta}-1}c_l\Rb_l$,
	where the matrix $\Rb_l$ is defined by:
 \vspace{-0.35cm}
	\begin{align*}
	\Rb_l(i,j)=\left\{
	\begin{array}{lr}
	1,~~&\text{if}~j=i+l,~~1\leq i\leq N,\\&\max(1,1+l)\leq j\leq \min(N+l,n_{\theta})\\
	0,&\text{otherwise.}
	\end{array}\right.
	\end{align*}
	Also, it is straightforward to show that
 \vspace{-0.3cm}
	\begin{equation}
	\text{vec}(\sum\limits_{l=-N+1}^{n_{\theta}-1}c_l\Rb_l) = \Wb\cb,
	\label{eq:subs1}
  \vspace{-0.3cm}
	\end{equation}
 \vspace{-0.2cm}
	in which
 \vspace{-0.15cm}
	\[
	\Wb=
	\begin{bmatrix}
	\text{vec}(\Rb_{-N+1})&\text{vec}(\Rb_{-N+2})&\cdots&\text{vec}(\Rb_{n_{\theta}-1})
	\end{bmatrix},
	\]
	\[
	\cb=
	\begin{bmatrix}
	c_{-N+1}&c_{-N+2}&\cdots&c_{n_{\theta}-1}
	\end{bmatrix}^T.
	\]
 \vspace{-0.2cm}
	Substituting \eqref{eq:subs1} in \eqref{eq:step3p} results in
 \vspace{-0.17cm}
	\begin{align}
	\nonumber(\cb_{k+1}, \Ub_{opt},\Zb_{opt}) =& \argmin_{\cb,\Ub\in\Uc,\Zb\in\Zc}\nt{\Wb\cb-\text{vec}(\Zb\Tb^{\frac{1}{2}})}^2\\&+\lambda^{\prime}\nt{\Wb\cb-\text{vec}(\Ub\Hb_{k+1})}^2.
	\label{eq:altrpro}
	\end{align}
	To solve \eqref{eq:altrpro}, we use cyclic minimization as follows:
	
	3.1) Update $\cb$:
 \vspace{-0.2cm}
	\begin{equation}
	\cb_{k+1} = \argmin_{\cb}\nf{\Wb\cb-\zb_{opt}^{\prime}}^2+\lambda^{\prime}\nt{\Wb\cb-\ub_{opt}}^2,
	\label{eq:altrpro1}
 \vspace{-0.2cm}
	\end{equation}
	in which $\zb^{\prime}_{opt}=\text{vec}(\Zb_{opt}\Tb^{\frac{1}{2}})$ and $\ub_{opt}=\text{vec}(\Ub_{opt}\Hb_{k+1})$. The first derivative of the above cost function should be zero. Hence, we have:
 \vspace{-0.3cm}
	\begin{equation}
	\cb_{k+1}=\frac{(\Wb^T\Wb)^{-1}}{1+\lambda^{\prime}}\Wb^T(\zb^{\prime}_{opt}+\lambda^{\prime}\ub_{opt}).
	\label{eq:tupdate}
	\end{equation}
	Using the above equation, the regressor matrix $\Phib$ can be updated as:
 \vspace{-0.35cm}
	\begin{equation}
	[c_{-N+1}^{k+1}~c_{-N+2}^{k+1}~\cdots~c_{n_{\theta}-1}^{k+1} ]=
	\cb_{k+1},\Phib_{k+1}=\sum_{l=-N+1}^{n_{\theta}-1} c_l^{k+1}\Rb_l.
 \label{eq:finphi}
 \vspace{-0.3cm}
	\end{equation}
 
	3.2) Update $\Zb$ and $\Ub$:
	
	In this step, we update the matrices $\Zb_{opt}$ and $\Ub_{opt}$ through the SVD, i.e.:
  \vspace{-0.2cm}
	\begin{align}
	\nonumber&(\Ib_F^d)^{\frac{1}{2}}\Phib_{k+1}^T=\Sb^{\prime}\Sigmab^{\prime}{\Ub^{\prime}}^T\rightarrow\Zb_{opt}={\Ub^{\prime}}{\Sb^{\prime}}^T\\
	&{\Hb_{k+1}\Phib_{k+1}^T}=\bar{ \Ub}\Sigb{\tilde{\Ub}}^T\rightarrow \Ub_{opt}={\tilde{\Ub}}\bar{\Ub}^T.
	\label{eq:svdupdate}
	\end{align}
	The pseudo-code of the final algorithm is summarized in Algorithm 1.
 \vspace{-0.3cm}
	\begin{algorithm}
		\caption{Low Coherence Input Design (LCID)-OMP}\label{euclid}
		\begin{algorithmic}
			
			\State \textbf{Input}: $\Ib_F^d$, $\Phib_1$, $\Hb_1$, $\varsigma$, $\lambda$, $\lambda^{\prime}$ and $0<c < 1$
			\For{$ k=1 $ to MaxIteration}\\
			$~~~\varsigma = c\varsigma$\\
			$~~~\Fb_k \triangleq \Hb_k^{T}\Hb_k-\Ib, \fb_{k}\triangleq\text{vec}(\Fb_k)$\\
			$~~~1) ~\Nb_{k+1}=\Fb_k-\text{vec}^{-1}(\Pc_{\Dc_2}(\fb_k))$ and update $\Nb_{k+1}^c$ by \\~~~~~~\eqref{eq:podema}\\
            $~~~~~\textbf{for }i=1\text{ to }3\textbf{ do}$\\
			$~~~2.1) ~\text{Update}~\Hb_{k+1}~\text{by \eqref{eq:udateh} and \eqref{eq:normali}}$\\
			$~~~2.2) ~\text{Update}~\Ub_{opt} ~\text{and}~\Qb_{opt} ~\text{by \eqref{eq:updateuq}}$\\
   $~~~~~\textbf{end for}$\\
$~~~~~\textbf{for }i=1\text{ to }3\textbf{ do}$\\
			$~~~3.1) ~\text{Update} ~~\Phib_{k+1}~~\text{by}~~ \eqref{eq:finphi}.$\\
			$~~~3.2) ~\text{Update}~\Zb_{opt} ~\text{and}~\Ub_{opt} ~\text{by \eqref{eq:svdupdate}}$\\
    $~~~~~\textbf{end for}$
			\EndFor
			\State \textbf{Output}: $\Phib_{k+1}$ and $\Hb_{k+1}$.
			\State $~~\xbh=((\Phib_{k+1}\Hb_{k+1}^{-1})^T(\Phib_{k+1}\Hb_{k+1}^{-1}))^{-1}(\Phib_{k+1}\Hb_{k+1}^{-1})^T\yb$
						\State \textbf{Support estimation:} ${\thetab^{\prime}}=\text{OMP}(\xbh,\Hb_{k+1},s)$, ${\Lambda}=\supp(\thetab^{\prime})$.
						\State \textbf{Sparse parameter estimation:} ${\thetab}=({\Phib_{k+1}}_{\Lambda_0})^{\dagger}\yb$
		\end{algorithmic}
		\label{alg:1}
	\end{algorithm}	
 \vspace{-0.4cm}
  { \begin{remark}
     We currently lack theoretical proof of convergence for the suggested approach, particularly concerning Step 3. However, the convergence of the proposed method was assessed experimentally through three different Monte Carlo studies (results not included in the paper due to space limitations), which demonstrated that the cost function in the optimization problem \eqref{eq:mainpr} decreases monotonically in the alternating minimization in Algorithm \ref{alg:1}.
For further insights and characteristics of alternating minimization, we also refer to \cite{TropDHS05, Powell1973,Stoica2008}. Additionally, it is worth noting that the proposed algorithm may also serve as a means to determine initial values when addressing \eqref{eq:conspro} with a general purpose constrained nonlinear optimization algorithm.
 \end{remark}}
  \vspace{-0.4cm}
	\section{NUMERICAL STUDY}\label{sec:nums}
 \vspace{-0.2cm}
	This section presents experimental evaluations of the proposed method in a model reference control problem studied in \cite{Hjalmarsson2009}. The proposed method can be utilized in
	conjunction with any sparse estimation algorithm, so that
	first the regressors $\Phib$ and $\Hb$ are designed through
	Algorithm \ref{alg:1}, and then the sparse estimation algorithm is applied
	to the model \eqref{eq:newmod}. In this study, we use the proposed method
	in conjunction with the OMP, LADMM \cite{BoydPCPE11} (Lasso via Alternating Direction Method of Multipliers) and BIC \cite{Ljung1999}, we denote these combinations by LCID-OMP, LCID-LADMM, and LCID-BIC, respectively.
	
	The simulations were performed on a laptop with a 4.00 GHz I7 CPU and 16GB RAM, using MATLAB R2022a and Microsoft Windows 10.
	
	The objective of model reference control problem is to design a feedback controller to achieve a sensitivity function that is as close as possible to the desired sensitivity $S$. The true system is assumed to be linear and time-invariant and given by
 \vspace{-0.35cm}
	\begin{equation}
	y(t)=G_0(q)u(t)+e(t),
	\label{eq:mplti}
 \vspace{-0.25cm}
	\end{equation}
The true system has the structure $G(q;\thetab)=\sum\limits_{k=1}^{n_{\theta}}{\theta}_k q^{-k}$ such that $G(q,\thetab_0)=G_o(q)$
and the exact order $n_\theta$ is unknown. {  An unknown model order can be viewed as a structured type of sparsity and here we will test if the proposed method is able to help also with the model order selection problem.} Two different values for $n_{\theta}$ are considered, namely $n_{\theta}=40$ and $n_{\theta}=30$. However, in both cases, only the first 10 coefficients of $\thetab$ are non-zero and the number of observations is $N=100$. 
Also, the noise ${e(t)}$ is modeled as a realization of an independent zero-mean Gaussian random variable with variance $\sigma^2$, which is chosen to obtain Signal-to-Noise Ratios (SNRs) ranging from 5 to 29. The SNR is defined as $\snr=10\log\frac{\nt{\Phib\thetab_0}^2}{\EX({\eb^T\eb})}$.

	A model reference controller when the given model $G$ is minimum phase, is
	as follows:
 \vspace{-0.35cm}
	\begin{equation}
	C(G)=\frac{1}{G}\frac{1-S}{S}.
	\label{eq:refco}
 \vspace{-0.25cm}
	\end{equation}
Here $S$ is the desired sensitivity function which we assume is given by
\vspace{-0.38cm}
	\begin{equation}
	S=\frac{1-q^{-1}}{1-\frac{1-\eta}{1+\eta}q^{-1}},~~~\eta\in[0,1],
 \vspace{-0.2cm}
	\end{equation}
	where $\eta$ controls the desired closed-loop bandwidth. Using the reference model controller \eqref{eq:refco} results in the following achieved sensitivity function
 \vspace{-0.35cm}
	\begin{equation}
	S(G) = \frac{1}{1+G_0C(G)}.
 \vspace{-0.2cm}
	\end{equation}
{\color{black} The study uses an application-oriented input design method to design the desired FIM and input spectrum. Application-oriented input design requires the performance degradation cost of the parameters estimate, which directly represents the performance of the application, to be less than a pre-specified bound $\frac{1}{2\gamma}$. One way to ensure this constraint is expressed in \cite{Hjalmarsson2009} as
\vspace{-0.3cm}
	\begin{equation}
	\Ib_F\geq\gamma\chi_\alpha^2(n_\theta) V^{''}_{app}(\thetab_0),
	\label{eq:LMI}
 \vspace{-0.15cm}
	\end{equation} 
 in which $V^{''}_{app}(\thetab_0)$ and $\chi_\alpha^2(n_\theta)$ denote the Hessian of the performance degradation cost and $\alpha$-percentile of the chi-square distribution with $n_\theta$ degrees of freedom, respectively.
 
	In this study, the following performance degradation cost  based on the $H_2$ error is considered
 \vspace{-0.2cm}
	\begin{equation}
	V_{app}(G)=\frac{1}{2}\Big\|\frac{S(G)-S}{S}\Big\|_2^2,
	\label{eq:redeg}
  \vspace{-0.2cm}
	\end{equation}
 so that the set of acceptable models is given by
 \vspace{-0.2cm}
 \begin{equation}
 \tag{47.b}
     \varepsilon_{app} =\{\thetab:V_{app}(\thetab)\leq \frac{1}{2\gamma}\}.
     \vspace{-0.1cm}
 \end{equation}
	The proposed method is evaluated using two performance measures: the performance degradation cost \eqref{eq:redeg} and the normalized mean square error (NRMSE) defined as NRMSE$=\frac{\nt{\thetab_0-{\thetab}}}{\nt{\thetab_0}}$, where $\thetab$ denotes the estimated parameters.}
	
 Additionally, the input spectrum is taken to have a finite representation with an order of $J$ given by
 \vspace{-0.25cm}
	\begin{equation}
	\Phi_u(\omega) = \sum_{k=-J+1}^{J-1}r_k e^{j\omega k},
 \vspace{-0.2cm}
	\end{equation}
	where $\{r_k\}$ denotes the autocorrelation coefficients of the input signal, i.e.,
	$r_k = \EX\{u(t)u(t-k)\}~~\text{and}~~r_k = r_{-k}.$

	The order of the input spectrum $J$ is set to 50 throughout this study.
 
	With the experimental measured by the used input power, the resulting application-oriented input design problem is given by
 \vspace{-0.45cm}
	\begin{align}
	\nonumber\min_{{\Phi_u(\omega)}} ~~~&\frac{1}{2\pi}\int_{-\pi}^{\pi}\Phi_u(\omega)\mathrm{d} \omega\\
	\nonumber\sbt~~ &\Phi_u(\omega)=\sum_{k=-J+1}^{J-1}r_k e^{j\omega k}\geq 0\\
	& 	\Ib_F\geq\gamma\chi_\alpha^2(n_\theta) V^{''}_{app}(\thetab_0),
	\label{eq:inpdes1}
	\end{align}
	where
  \vspace{-0.2cm}
	\begin{align}
	\nonumber&V_{app}^{''}(\thetab) = \frac{1}{2\pi}\int_{-\pi}^{\pi}\Big|\frac{1-S(e^{j\omega})}{G_0(e^{j\omega})}\Big|^2G^{'}(\theta_0,e^{j\omega})(G^{'}(\theta_0,e^{j\omega}))^{*}d\omega.
	\label{eq:hessian}
  \vspace{-0.2cm}
	\end{align}
	The above Hessian matrix is computed by numerical differentiation in MATLAB.
	
	To solve the optimization problem \eqref{eq:inpdes1}, the MOOSE toolbox\footnote{For MOOSE2, we used the MATLAB code available at `\url{https://www.kth.se/polopoly_fs/1.195834.1600688635!/Menu/general/column-content/attachment/moose_ver01f.zip}'. }\cite{Annergren2016} is utilized. The toolbox provides the corresponding optimal FIM, denoted as $\Ib_F^d$. In the FDM, the input is generated by filtering white noise through the stable minimum phase spectral factor of the obtained input spectrum.

	In this study, hyperparameters are used in all algorithms and are tuned using Cross Validation (CV). The data is split into a training set (80$\%$ of the data) and a validation set (20$\%$ of the data). Initially, the parameter vector $\thetab$ is estimated using the training set based on 200 different hyperparameters that are logarithmically spaced between $10^{-5}$ and $10$. Subsequently, the hyperparameters that result in the minimum error $\nt{\yb_v-\Phib_v\hat{\thetab}}$ are selected using the validation set, where $\yb_v$ and $\Phib_v$ indicate the observations and regressor matrix for the validation set, respectively. To obtain the minimum error, a sufficiently small and large range of hyperparameters, i.e., $10^{-5}$ and $10$, respectively, are chosen. {\color{black} The hyperparameter of the algorithms are: $\lambda$ (LADMM), $\lambda$ and $\lambda^{\prime}$ (proposed method), and $\lambda$ (CORED-OMP as in \cite{10195960}). 
	
	 Other parameters of the proposed algorithms (see Algorithm \ref{euclid})  are set as $\varsigma = 1$,  and $c_1 = 0.97$.}
	
	{\color{black} In the simulations, we compare the performance of the proposed methods, i.e., LCID-OMP, LCID-LADMM, and LCID-BIC with standard least-squares (LS) used in conjunction with AICc (small sample-size corrected AIC) and BIC to select the model order, resulting in LS-AICc and LS-BIC, respectively. Additionally, the comparison includes OMP, LADMM, the method proposed in \cite{9838242} referred to as ARCID-OMP (Approximating the Reduction of Coherence in Input Design via OMP), and CORED-OMP \cite{10195960}. In LS-AICc, LS-BIC, OMP, and LADMM, the input is designed using the FDM explained above.} In CORED-OMP, the parameters of the system are estimated using the coordinate transformation \eqref{eq:mainpr1} in which the regressor $\Phib$ is known from \eqref{eq:inf1}.
	
	Additionally, we benchmark the performance with an oracle knowing the sparsity pattern of $\thetab_0$. This method denoted FDM-Known Sparsity, estimates the non-zero parameters using standard least-squares with the input obtained from \eqref{eq:inpdes1}. 
	
	In order to assess the performance of the different methods, 400 experiments are conducted per $\text{SNR} = [5~8~11~14~17~20~23~26~29]$. The mean values of NRMSE and $V_{app}(\thetab)$ are then computed for each method. It is worth noting that although the different methods may result in varying input powers, the same SNRs are used for each method for a fair comparison.

{\color{black} The mutual coherence of the regressor $\Phib$ in this study, denoted by $\mu_{\Phib}$, is 0.98 which is very high and fails to meet the required upper bound in Theorem \ref{thm:l1normsp}, whereas the mutual coherence of the regressor in the new coordinates, represented by $\mu_{\Hb}$ for CORED-OMP and LCID, is 0.05. Additionally, the obtained $\mu_{\Hb}$ for ARCID-OMP is 0.73. Therefore, this numerical study demonstrates that while the optimal regressor $\Phib$ exhibits a high mutual coherence, we can obtain a small mutual coherence through the proposed method by using the coordinate transformation matrix $\Hb$.}
	
\begin{figure*}[ht!]
   \begin{minipage}{0.28\textwidth}
    \centering	\centerline{\includegraphics[width=0.9\columnwidth]{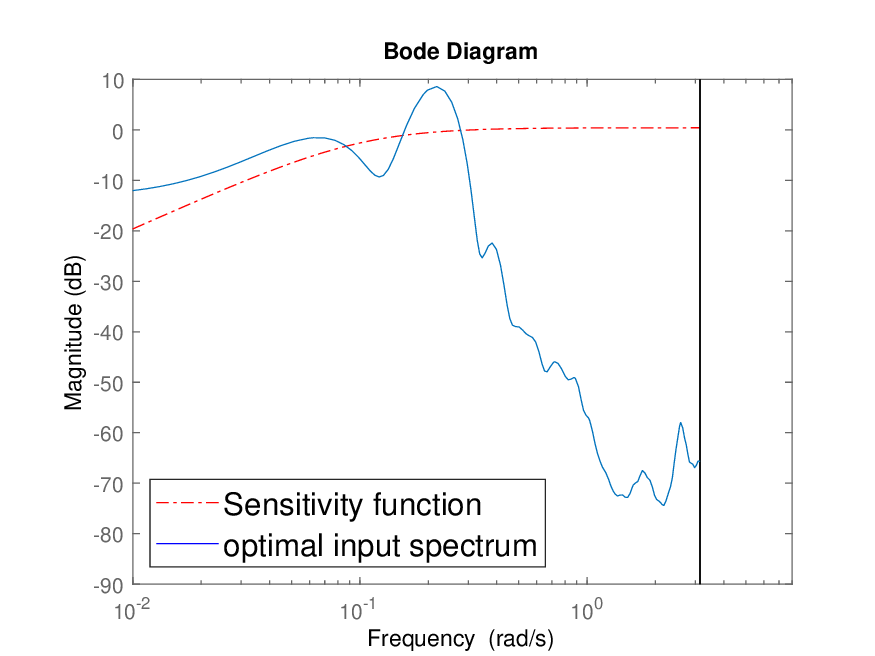}}
   \vspace{-0.15cm}
		\caption{Evaluation with assumption: $ \Phib\in\Rbb^{100\times 40}$,
			$s = 10$ and $\zeta = 0.1$.}
		\label{fig:sn3r}
   \end{minipage}\hspace{\fill}
   \begin{minipage}{0.28\textwidth}
    \centering
		\centerline{\includegraphics[width=0.9\columnwidth]{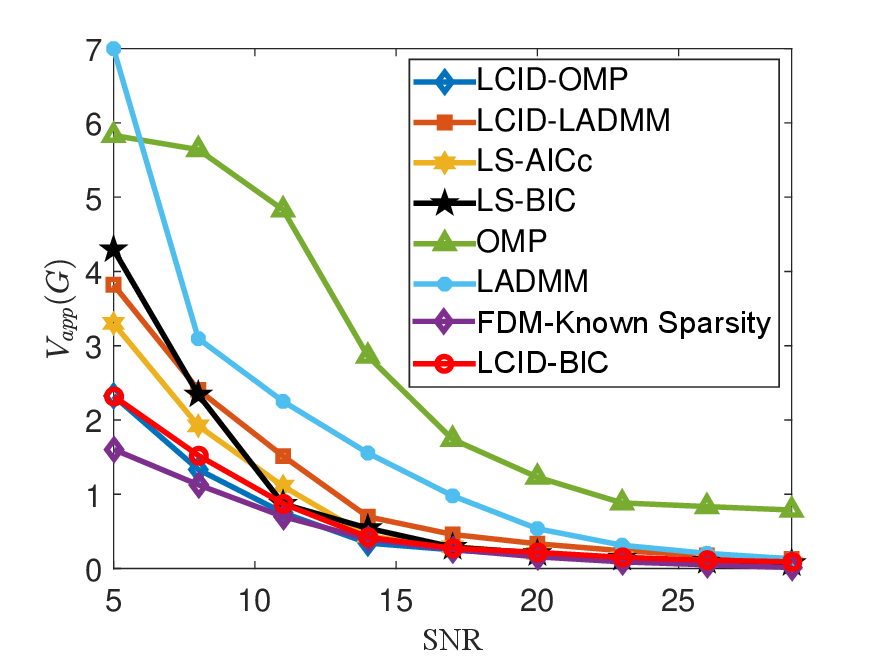}}
  \vspace{-0.15cm}
		\caption{Evaluation of performance degradation with assumptions: $ \Phib\in\Rbb^{100\times 40}$ and
			$s = 10$.}
		\label{fig:snr4}
   \end{minipage}\hspace{\fill}
    \begin{minipage}{0.28\textwidth}
   \centering
		\centerline{\includegraphics[width=0.9\columnwidth]{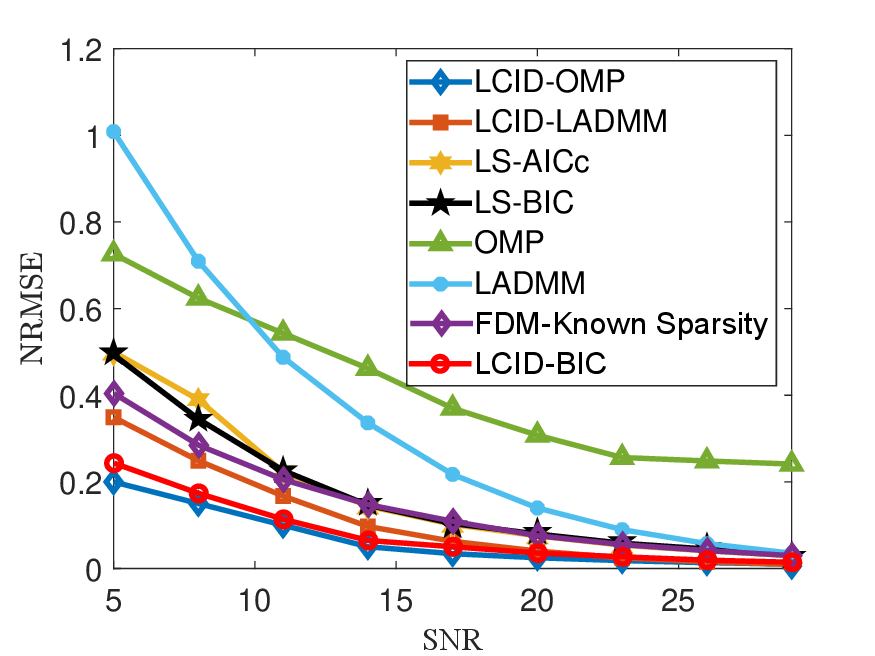}}
  \vspace{-0.1cm}
		\caption{Evaluation of NRMSE with assumptions: $ \Phib\in\Rbb^{100\times 40}$ and
			$s = 10$.}
		\label{fig:snr3}
   \end{minipage}\hspace{\fill}
   \vspace{-0.5cm}
\end{figure*}

{\color{black}	Figure \ref{fig:sn3r} displays the ideal input spectrum and sensitivity function for $\Phib\in\Rbb^{100\times 40}$, where $s=10$ and $\eta=0.1$ obtained by solving \eqref{eq:inpdes1}. This figure indicates that the crucial frequency interval of the optimal input spectrum, i.e., $[0.1, 0.3]$rad/s, coincides with the cross-over frequency interval of the desired sensitivity function. Furthermore, the optimal input spectrum is negligible at other frequencies. The FDM thus generates a highly correlated input sequence.}
	
	{\color{black} The results of performance degradation cost and NRMSE for $\Phib\in\Rbb^{100\times 40}$ and $s = 10$ are reported in Figures \ref{fig:snr4} and \ref{fig:snr3}, respectively. In Figure \ref{fig:snr4}, we see that FDM with known sparsity performs the best. This is of course expected since the optimal input design is used together with known sparsity pattern. However, significant performance improvements are observed when the proposed method is utilized with OMP, LADMM and BIC. As a result, LCID-OMP and LCID-BIC outperform LS-AICc and LS-BIC significantly to reduce the model order, for low SNRs.
	Turning to Figure \ref{fig:snr3}, we see, perhaps surprisingly, that the three new methods (LCID-OMP, LCID-LADMM, and LCID-BIC) have lower NRMSE than the oracle method  FDM-Known Sparsity. This is because in FDM-Known Sparsity, although the sparsity pattern is known, the obtained optimal regressors are highly correlated, making it difficult for an estimator to distinguish between different regressors. On the other hand, the mutual coherence in LCID-OMP is low, resulting in better estimation performance. We also see that the three new methods outperform all other methods, including LS-AICc and LS-BIC. Therefore, this study highlights the efficacy of the proposed method in improving estimation accuracy and selecting model order.}
	
\begin{figure*}[ht!]
   \begin{minipage}{0.28\textwidth}
     \centering

\centerline{\includegraphics[width=0.9\columnwidth]{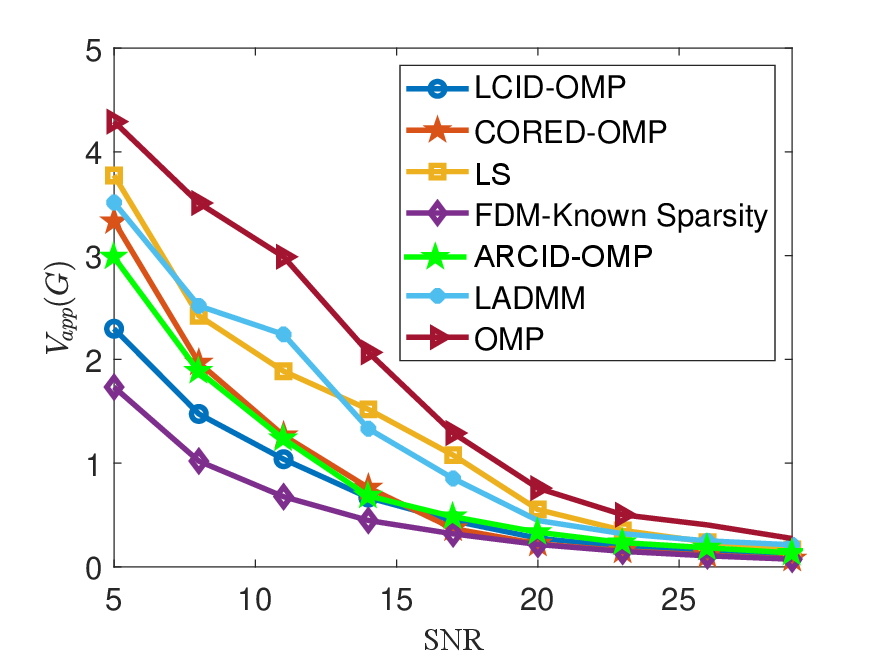}}
   \vspace{-0.1cm}
		\caption{Evaluation of performance degradation with assumptions: $ \Phib\in\Rbb^{100\times 30}$ and
			$s = 10$.}
		\label{fig:snr6}
   \end{minipage}
   \hspace{\fill}
   \begin{minipage}{0.28\textwidth}
    \centering
		\centerline{\includegraphics[width=0.9\columnwidth]{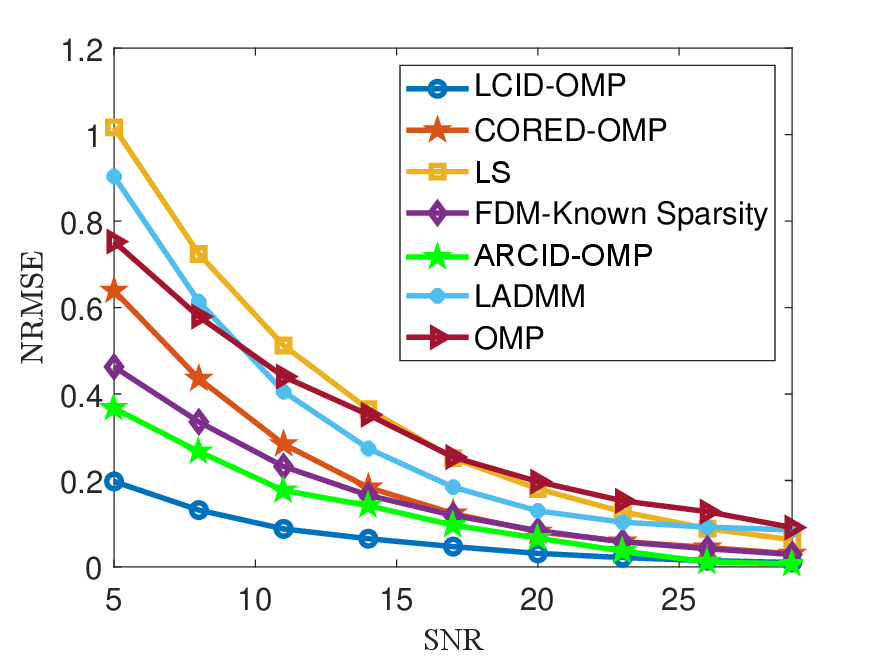}}
  \vspace{-0.1cm}
		\caption{Evaluation of NRMSE with assumptions: $ \Phib\in\Rbb^{100\times 30}$ and
			$s = 10$.}
		\label{fig:snr5}
   \end{minipage}
      \hspace{\fill}
\begin{minipage}{0.26\textwidth}
	\centering
     \begin{adjustbox}{center, width=0.85\columnwidth}
     \begin{tabular}{|c|c|}
			\hline 
			\text{ Algorithm }&\text{ $\snr=15$ }\tabularnewline
			\hline 
			\hline 
			\multirow{1}{*}{\text{LCID-OMP}} &{\text{18.9s}}
			\tabularnewline
			\cline{1-2} 
			\multirow{1}{*}{\text{CORED-OMP}}  &{\text{12.1s}}
			\tabularnewline
			\cline{1-2}
			\multirow{1}{*}{\text{LS}} &{\text{9.1s}}
			\tabularnewline
			\cline{1-2}
			\multirow{1}{*}{\text{FDM-Known Sparsity}} &{\text{7.53s}}  
			\tabularnewline
			\cline{1-2} 
			\multirow{1}{*}{\text{OMP}} &{\text{9.15s}}  
			\tabularnewline
			\cline{1-2}
			\multirow{1}{*}{\text{LADMM}} &{\text{9.12s}} 
			\tabularnewline
			\cline{1-2}
			\multirow{1}{*}{\text{LCID-LADMM}} &{\text{19s}} 
   \tabularnewline
			\cline{1-2}
			\multirow{1}{*}{\text{{ACRID-OMP}}} &{\text{{548s}}} 
			\tabularnewline
			\cline{1-2} 
			\hline 
		\end{tabular}
   \end{adjustbox}
   \vspace{-0.1cm}
\captionof{table}{Average running times (in seconds) to design optimal input with assumptions:$ \Phib\in\Rbb^{100\times 30}$ and
			$s = 10$.}
		\label{tbl:srr}
   \end{minipage}
   \vspace{-0.85cm}
\end{figure*}

	{\color{black} Figures \ref{fig:snr6} and \ref{fig:snr5} depict the results of performance degradation cost and NRMSE, respectively, for $\Phib\in\Rbb^{100\times 30}$ and $s = 10$. In this study, the LCID-OMP outperforms CORED-OMP significantly. A noteworthy observation from these figures is that the proposed method yields lower performance degradation cost and NRMSE compared to ARCID-OMP. This difference can be attributed to the exact formulation of mutual coherence within the proposed method, resulting in a calculated $\mu_{\Hb}$ of 0.05 for LCID, in contrast to the value of 0.73 obtained for ARCID. 
	This numerical analysis thus provides compelling evidence for the substantial improvement achieved by LCID over ARCID.}
 
	According to the all figures, the proposed method has a significant effect on both OMP and LADMM, so that using LCID greatly reduces the NRMSE and $V_{app}(\thetab)$.
	
	{\color{black}To measure the complexity of the methods, we report computational time which is required to design the optimal regressor and estimate the parameters of model for $\Phib\in\Rbb^{100\times 30}$, $s = 10$ and $\snr = 15$ in Table \ref{tbl:srr}. As shown in this table, the computational requirement is higher using LCID than the other methods, except for ARCID, still the computational overhead is in no way dramatic. An intriguing observation from this table is that utilizing the closed-form solution in LCID, as opposed to employing the CVX toolbox in ARCID, leads to a significant reduction in computational time.}
	
	The proposed method involves designing the regressor and coordinate transformation jointly, whereas in CORED-OMP, the optimal regressor design is independent of the coordinate transformation. Therefore, in LCID-OMP and CORED-OMP, a regressor with low mutual coherence is obtained in the new coordinates. However, compared to CORED-OMP, the coordinate transformation in LCID-OMP has a lower undesired effect on the noise. This is why LCID-OMP exhibits a better ability to estimate the sparse parameter vector than other methods. Additionally, the error $\frac{\nf{\Phib^T\Phib-\Ib_F^d}}{\sqrt{N n_{\theta}}}$ for LCID and other methods when $n_{\theta} = 40$ is 0.018 and 0.0001, respectively, which means the proposed method sacrifices the error $\frac{\nf{\Phib^T\Phib-\Ib_F^d}}{\sqrt{N n_{\theta}}}$ slightly to achieve higher accuracy in estimating the parameters. 
 \vspace{-0.15cm}
 \vspace{-0.35cm}
	\section{Conclusion} 
 \vspace{-0.2cm}
{\color{black}	The paper presents an algorithm for designing optimal inputs in models with a sparse parameter vector. The method is applicable to various commonly used models in system identification, such as FIR, Laugerre, Kautz, and generalized basis function expansion
models. The proposed method is based on minimizing the difference between the  Fisher information matrix and the desired Fisher matrix, along with a coordinate transformation.

To compute the optimal input, we utilize cyclic minimization involving three sub-steps and proximal mapping.

We have also applied the method to a model reference control problem where the actual model order is unknown. Here, the proposed method, LCID, combined with OMP, LADMM, and BIC, significantly improves the accuracy of estimating the sparse parameter vector compared to other algorithms. When there is no prior knowledge about the location of non-zero entries in the parameter vector, LCID-OMP, and LCID-ADMM have lower performance degradation costs than other methods. The numerical study also showed that model order selection criteria such as AICc and BIC benefit significantly from the method for low SNRs.

The proposed method can be extended to incorporate input design for nonlinear systems, where the regressor $\Phib$ includes the nonlinear components of input sequences. It is noteworthy that many nonlinear models, including the Volterra series and non-linear Auto-Regressive models with eXogenous inputs (NARX), are inherently sparse. This important topic is under current research.} 
\vspace{-0.7cm}
	\bibliography{myref}

\begin{thebibliography}{10}
\providecommand{\url}[1]{#1}
\csname url@samestyle\endcsname
\providecommand{\newblock}{\relax}
\providecommand{\bibinfo}[2]{#2}
\providecommand{\BIBentrySTDinterwordspacing}{\spaceskip=0pt\relax}
\providecommand{\BIBentryALTinterwordstretchfactor}{4}
\providecommand{\BIBentryALTinterwordspacing}{\spaceskip=\fontdimen2\font plus
\BIBentryALTinterwordstretchfactor\fontdimen3\font minus \fontdimen4\font\relax}
\providecommand{\BIBforeignlanguage}[2]{{%
\expandafter\ifx\csname l@#1\endcsname\relax
\typeout{** WARNING: IEEEtran.bst: No hyphenation pattern has been}%
\typeout{** loaded for the language `#1'. Using the pattern for}%
\typeout{** the default language instead.}%
\else
\language=\csname l@#1\endcsname
\fi
#2}}
\providecommand{\BIBdecl}{\relax}
\BIBdecl

\bibitem{Ljung1999}
L.~Ljung, \emph{System Identification: Theory for the User}, 2nd, Ed.\hskip 1em plus 0.5em minus 0.4em\relax Prentice-Hal, 1999.

\bibitem{293196}
B.~Wahlberg, ``System identification using kautz models,'' \emph{IEEE Transactions on Automatic Control}, vol.~39, no.~6, pp. 1276--1282, 1994.

\bibitem{Heuberger&Hof&Bogsra:95}
P.~Heuberger, {P.M.J.\ {Van den Hof}}, and O.~Bosgra, ``A generalized orthonormal basis for linear dynamical systems,'' \emph{IEEE Transactions on Automatic Control}, vol.~40, no.~3, pp. 451--465, March 1995.

\bibitem{RePEc}
J.~Kennington, E.~Olinick, and D.~Rajan, Eds., \emph{{Wireless Network Design}}.\hskip 1em plus 0.5em minus 0.4em\relax Springer, June 2011, no. 978-1-4419-6111-2.

\bibitem{6034080}
A.~P. Kannu and P.~Schniter, ``On communication over unknown sparse frequency-selective block-fading channels,'' \emph{IEEE Transactions on Information Theory}, vol.~57, no.~10, pp. 6619--6632, 2011.

\bibitem{10195960}
J.~Parsa, C.~R. Rojas, and H.~Hjalmarsson, ``Transformation of regressors for low coherent sparse system identification,'' \emph{IEEE Transactions on Automatic Control}, pp. 1--16, 2023.

\bibitem{Chiuso2012}
A.~Chiuso and G.~Pillonetto, ``A bayesian approach to sparse dynamic network identification,'' \emph{Automatica}, vol.~48, no.~8, pp. 1553--1565, 2012.

\bibitem{Elad10}
M.~Elad, \emph{Sparse and Redundant Representations}.\hskip 1em plus 0.5em minus 0.4em\relax Springer, 2010.

\bibitem{Ben-Haim2010}
Z.~Ben-Haim, Y.~C. Eldar, and M.~Elad, ``Coherence-based performance guarantees for estimating a sparse vector under random noise,'' \emph{{IEEE} Transactions on Signal Processing}, vol.~58, no.~10, pp. 5030--5043, oct 2010.

\bibitem{8553417}
J.~Parsa, M.~Sadeghi, M.~Babaie-Zadeh, and C.~Jutten, ``Joint low mutual and average coherence dictionary learning,'' in \emph{2018 26th European Signal Processing Conference (EUSIPCO)}, 2018, pp. 1725--1729.

\bibitem{9838242}
J.~Parsa, C.~R. Rojas, and H.~Hjalmarsson, ``Optimal input design for sparse system identification,'' in \emph{European Control Conference (ECC)}, 2022, pp. 1999--2004.

\bibitem{PatiRK93}
Y.~C. Pati, R.~Rezaiifar, and P.~S. Krishnaprasad, ``Orthogonal matching pursuit: recursive function approximation with applications to wavelet decomposition,'' in \emph{In Proc. Asilomar Conf. Signal Syst. Comput.}, 1993.

\bibitem{WangKwonShim2012}
J.~Wang, S.~Kwon, and B.~Shim, ``Generalized orthogonal matching pursuit,'' \emph{IEEE Trans. on Signal Proc.}, vol.~60, no.~12, pp. 6202--6216, 2012.

\bibitem{Goodwin1977}
G.~Goodwin and R.~Payne, \emph{Dynamic System Identification: Experiment Design and Data Analysis}.\hskip 1em plus 0.5em minus 0.4em\relax Academic Press, New York, 1977.

\bibitem{Bombois2006}
X.~Bombois, G.~Scorletti, M.~Gevers, P.~Van~den Hof, and R.~Hildebrand, ``Least costly identification experiment for control,'' \emph{Automatica}, pp. 1651--1662, 2006.

\bibitem{Hjalmarsson2009}
H.~Hjalmarsson, ``System identification of complex and structured systems,'' \emph{European Journal of Control}, pp. 275--310, 2009.

\bibitem{EBADAT20141422}
A.~Ebadat, B.~Wahlberg, H.~Hjalmarsson, C.~Rojas, P.~Hägg, and C.~Larsson, ``Applications oriented input design in time-domain through cyclic methods,'' \emph{IFAC Proceedings Volumes}, vol.~47, no.~3, pp. 1422--1427, 2014, 19th IFAC World Congress.

\bibitem{PariB14}
N.~Parikh and S.~Boyd, ``Proximal algorithms,'' \emph{Foundations and Trends in Optimization}, vol.~1, no.~3, pp. 123--231, 2014.

\bibitem{Stoica2008}
P.~Stoica, J.~Li, X.~Zhu, and B.~Guo, ``Waveform synthesis for diversity-based transmit beampattern design,'' \emph{IEEE Transactions on Signal Processing}, vol.~56, pp. 2593--2598, 2008.

\bibitem{TropDHS05}
J.~A. Tropp, I.~S. Dhillon, R.~W. Heath, and T.~Strohmer, ``Designing structured tight frames via an alternating projection method,'' \emph{IEEE Trans. Info. Theory}, vol.~51, no.~1, pp. 188--209, 2005.

\bibitem{Powell1973}
M.~J.~D. Powell, ``On search directions for minimization algorithms,'' \emph{Mathematical Programming}, vol.~4, no.~1, pp. 193--201, 1973.

\bibitem{BoydPCPE11}
S.~Boyd, N.~Parikh, E.~Chu, B.~Peleato, and J.~Eckstein, ``Distributed optimization and statistical learning via the alternating direction method of multipliers,'' \emph{Foundations and Trends in Machine Learning}, vol.~3, no.~1, pp. 1--122, 2011.

\bibitem{Annergren2016}
M.~Annergren and C.~Larsson, ``{MOOSE}2{\textemdash}a toolbox for least-costly application-oriented input design,'' \emph{{SoftwareX}}, vol.~5, pp. 96--100, 2016.

\end{thebibliography}
	\bibliographystyle{IEEEtran}
\end{document}